\documentclass[prb,preprint,showpacs,preprintnumbers,amsmath,amssymb]{revtex4}

\usepackage{graphicx}
\usepackage{dcolumn}
\usepackage{bm}
\usepackage{textcomp}

\begin{document}

\preprint{APS/123-QED}

\title{Lateral Signals in Piezoresponse Force Microscopy at Domain Boundaries of Ferroelectric Crystals}

\author{Florian Johann\footnote{Present address: Max Planck Institute for Microstructure Physics, Weinberg 2, 06120 Halle, Germany}}
\author{Tobias Jungk}
\author{Martin Lilienblum}
\author{\'{A}kos Hoffmann}
\author{Elisabeth Soergel}
\email{soergel@uni-bonn.de}

\affiliation{Institute of Physics, University of Bonn,
Wegelerstra\ss e 8, 53115 Bonn, Germany}


\begin{abstract}
In piezoresponse force microscopy a lateral signal at the domain boundaries is occasionally observed. In recent years, a couple of experiments have been reported and varying explanations for the origin of this lateral signal have been proposed. Additionally, elaborated theoretical modeling for this particular issue has been carried out. Here we present experimental data obtained on different crystallographic cuts of $\rm LiNbO_3$, $\rm BaTiO_3$, and $\rm KTiOPO_4$ single crystals. We could thereby rule out some of the explanations proposed so far, introduce another possible mechanism, and quantitatively compare our results to the existing modeling.
\end{abstract}

\pacs{77.65.-j, 68.37.Ps, 77.80.Dj}

\maketitle

Piezoresponse force microscopy (PFM) has become a standard technique for the investigation of ferroelectric domain patterns during the past fifteen years.\cite{Gue92,Alexe} For PFM a scanning force microscope is operated in contact mode with an oscillating voltage applied to the tip. The piezomechanical response of the sample, i.\,e., the vibration of the sample surface beneath the tip, is read out via the movement of the cantilever using a lock-in amplifier. When speaking about PFM imaging of ferroelectric domains, one usually refers to measurements recording the deflection of the cantilever ('vertical signal')  owing to an out-of plane driving force caused by the thickness changes of the sample.

In scanning force microscopy, however, it is also possible to record in-plane driving forces via the torsion of the cantilever ('lateral signal'). A lateral signal observed by PFM indicates an in-plane piezomechanical deformation of the sample surface.\cite{Eng99,Kal06}  In case of a contrast detected between different domain faces this could be attributed to an expansion/contraction mechanism (rather than shearing) caused by the in-plane electric field components arising from the PFM tip.\cite{Jun09} But also at the domain boundaries a distinct lateral signal was occasionally observed.~\cite{Gan02,Wit02,Scr05,Jun06b,Guy09} The origin of this lateral signal at the domain boundaries (LS$_{\rm db}$) is still under discussion. In this contribution we review the mechanisms proposed, compare them to our experimental results, and outline the complexity of the situation.


The first publications showing a LS$_{\rm db}$ supposed the topographical slope at the domain boundary to cause the torsion of the cantilever.~\cite{Gan02,Wit02,Scr05} This slope originates from the opposed piezoelectric deformation of the crystal left and right from the tip when it is located on top of a domain boundary. To determine the angle $\alpha$ of the slope, the appropriate piezoelectric tensor element of the crystal $d_{ij}$ and the voltage $U$ applied to the tip is needed.  Using the simple assumption (SA) that the deformation of the surface takes place on the dimension of the tip diameter~\cite{Jun06b} or by finite element method (FEM)\cite{Scr05,David} yields $\alpha_{\rm SA}=0.04^{\circ}$ and $\alpha_{\rm FEM}=0.0062^{\circ}$ respectively, for lithium niobate ($\rm LiNbO_3$) with $d_{33}=7$\,pm/V and \mbox{$U=\rm 10\,V$}.\cite{Oka89} Unfortunately, in the publications assuming the slope to cause LS$_{\rm db}$ neither a detailed description of the suggested mechanisms nor a quantitative estimate of the expected LS$_{\rm db}$ is given. The following considerations are therefore based on the figures published in those papers.

In Ref.~[\onlinecite{Gan02}] the authors state that differential piezoelectric activity in adjacent domains gives rise to a torque of the cantilever. Neither the schematic shown (original Fig.~6(c)) nor the text itself, however, does allow to draw conclusions on their understanding of the exact mechanism leading to torsion.

\begin{figure}[hhh]
\includegraphics{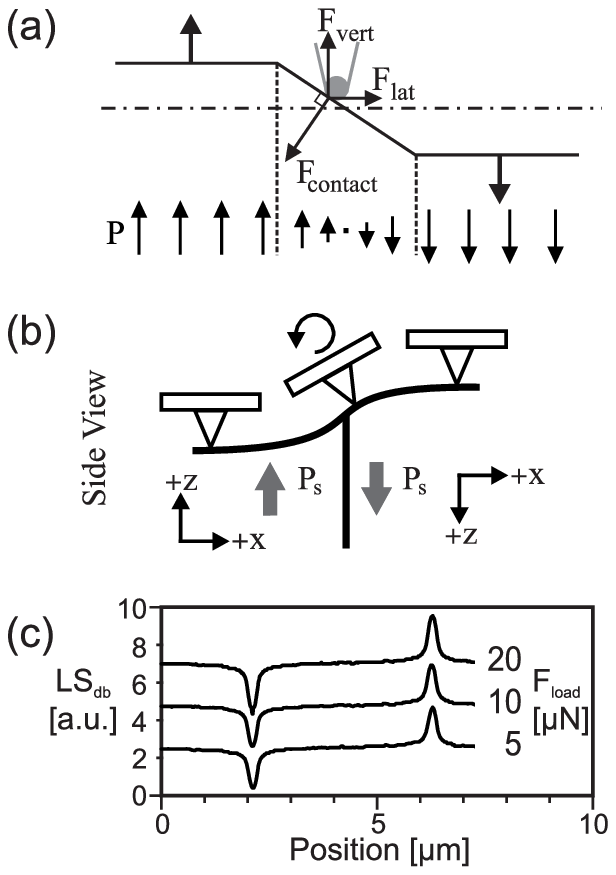}
\caption{\label{fig:Joh-01}
Two mechanisms for the lateral signal at the domain boundaries relying on the topographical slope caused by the opposed piezoelectric deformation of the crystal beneath the tip.
(a)~Drawing adopted from Fig.~2 in Ref.~[\onlinecite{Wit02}]  and
(b)~and from Fig.~5(d) in Ref~[\onlinecite{Scr05}].
(c) Measurement of the $\rm LS_{bd}$ for different loading forces on a PPLN crystal across two domains boundaries using a tip with a spring constant of 22.7\,N/m.
}
\end{figure}

In Ref.~[\onlinecite{Wit02}] no friction between tip and sample surface seems to be assumed the contact force $F_{\rm contact}$ being drawn perpendicular to the inclined plane (Fig.~\ref{fig:Joh-01}(a)). Thus, the tip slides along the sample surface and the cantilever torques until equilibrium of forces is reached. Using the parallelogram of forces one can calculate the lateral force to be \mbox{$F_{\rm lat} = \tan \alpha \, F_{\rm load}$}, where $F_{\rm load}$ is the load of the tip (typically of the order of \mbox{10\,nN\,\cite{Col}}). This leads to lateral forces of $F_{\rm lat}\approx1\,{\rm pN}(\alpha_{\rm FEM}$) and  $7\,{\rm pN}(\alpha_{\rm SA})$.

In Ref.~[\onlinecite{Scr05}] strong friction between tip and sample surface seems to be assumed since the torsion of the cantilever is drawn as if the tip was taken along by the surface (Fig.~\ref{fig:Joh-01}(b)). The tip sticks to the sample surface and the cantilever torques until equilibrium of forces is reached. For simplicity we assume the extreme case that no sliding at all occurs. The torque experienced at the apex of the tip, approximated by a sphere (radius $r=60$\,nm), is $M_{\rm s}= r\,\sin\alpha\,F_{\rm load}$. This torque $M_s$ has to be balanced by the torsion of the cantilever leading to a torque $M_{\rm c}= k_{\rm tor} \, \Delta x  \, h$, where $k_{\rm tor}$\,[N/m]  is the torsion spring constant of the cantilever, $\Delta x$ the lateral displacement of the sphere, and $h \approx 10\,$\textmu m the tip height. The lateral force can thus be calculated as $F_{\rm lat} =k_{\rm tor} \, \Delta x =  {M_{\rm s}} / {h} =({r\,\sin\alpha} / {h}) \,F_{\rm load}$. Inserting the numbers gives $F_{\rm lat}\approx 6.5\,{\rm fN}(\alpha_{\rm FEM})$ and $42\,{\rm fN}(\alpha_{\rm SA})$.


As shown above, slope-based explanations for the LS$_{\rm db}$ require $F_{\rm lat} \propto F_{\rm load}$.
We therefore performed experiments on $\rm LiNbO_{3}$ where we changed $F_{\rm load}$ by two orders of magnitude (thereby utilizing probes with different spring constants), however, no effect on LS$_{\rm db}$ was observed. Figure~\ref{fig:Joh-01}(c) shows part of those measurements where during data acquisition, we changed the load of the tip (spring constant $k=22.7$\,N/m) by a factor of 4 by setpoint adjustment. In addition, we also performed measurements on $\rm LiNbO_{3}$ at different relative humidities (between 5\% and 50\%) thereby changing the meniscus force between tip and sample surface, which might act as an offset to the load of the tip, but again no effect on LS$_{\rm db}$ was observed. Our quantitative measurements on $\rm LiNbO_3$ gave values of $F_{\rm lat}\approx 10\,\rm nN$. This is much larger than the values estimated above. From these results we conclude that the LS$_{\rm db}$ can not be caused by the slope at the domain  boundary.

A non-mechanical model for the LS$_{\rm db}$ was proposed next, assuming the strong electric fields caused by the domain-specific polarization surface charging to account for the lateral signal at the domain boundaries.\cite{Jun06b} Experiments performed on
70\,nm thin PZT films, however, showed no significant contribution of surface charging to the LS$_{\rm db}$.\cite{Guy09} We now re-explored this issue, explicitly investigating the $x$- and $y$-faces of $\rm LiNbO_3$ and the $y$-face of $\rm KTiOPO_4$ crystals. All these faces being non-polar, no surface charging is expected, which we confirmed by electrostatic force microscopy~\cite{Ter89} measurements. In our PFM measurements of those samples, a  LS$_{\rm db}$ was observed on the $y$-face of LiNbO$_3$ only. In addition, we investigated the polar $z$-faces of $\rm KTiOPO_4$, LiNbO$_3$, and  3\%~Fe:LiNbO$_3$, which all showed a LS$_{\rm db}$. The LS$_{\rm db}$ on the iron doped LiNbO$_3$ was comparable to that on the undoped crystal. From these experiments we conclude that domain specific surface polarization charging can not be the only driving mechanism for the LS$_{\rm db}$.

Recently a theoretical model for the LS$_{\rm db}$ has been reported.~\cite{Mor07} In this work, the displacement of the sample surface underneath the tip is calculated, especially for scanning across 180$^{\circ}$ domain boundaries. Here the tip is assumed to follow this displacement. In the model, the LS$_{\rm db}$ is represented by an effective piezoelectric tensor element~$d_{35}^{\rm eff}$ which itself is derived from the non-zero tensor elements $d_{31}$, $d_{33}$ and $d_{15}$.
A very recent publication investigated 70\,nm thin PZT films and attributed the  LS$_{\rm db}$ to shear effects according to this  model.~\cite{Guy09} In this work, however, only qualitative measurements of the  LS$_{\rm db}$ are shown. A thorough comparison with the model was therefore not possible.

\begin{figure}[hhh]
\includegraphics{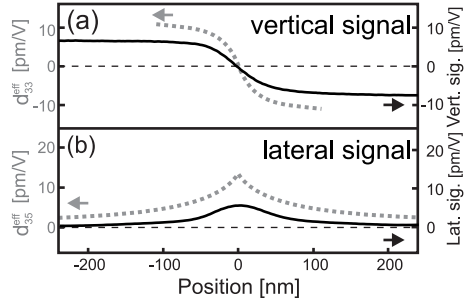}
\caption{\label{fig:Joh-02}
Comparison of calculated scanlines (dotted grey) [after Fig.~6(a) and (b) from Ref.~[\onlinecite{Mor07}]]
and measured data (black) obtained across a 180$^{\circ}$ domain wall at position $0$ on a LiNbO$_3$ crystal using a tip of 60\,nm radius.
}
\end{figure}

To check the validity of the model we performed quantitative measurements of the vertical and the lateral PFM signals using a periodically poled LiNbO$_3$ (PPLN) crystal. We calibrated our scanning force microscope with a setup built on purpose. Therefore we glued an aluminium cube on top of a  piezo slab. To one of the cube's upright faces we glued another piezo slab. As a sample, we attached a PPLN directly to the upright piezo slab. This setup allows for calibration and subsequent measurements  without the need of repeated approaches of the tip onto different samples. The error of the calibration was found to be approx.~10\%. Figure~\ref{fig:Joh-02} shows the calculated line scans for $\rm LiNbO_3$ from Ref.~[\onlinecite{Mor07}] in comparison with our experimental results obtained with a diamond coated tip of radius \mbox{$r= 60\,\rm nm$} (confirmed by scanning electron microscopy). Neither the amplitudes nor the shapes are well reproduced by the theoretical calculations.

\begin{figure}[hhh]
\includegraphics{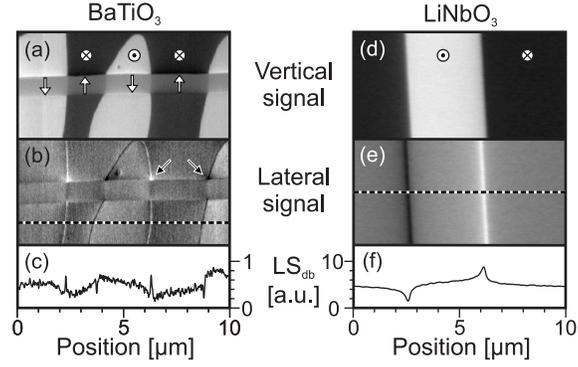}
\caption{\label{fig:Joh-03}
Measurements on multi-domain single crystals of BaTiO$_3$ and LiNbO$_3$. The vertical and lateral PFM signals (a, b) and (d, e) were recorded simultaneously. The horizontal grey stripe in (a) and (b) shows $a$-domains, their orientation is given by the arrows.  The amplitude of the lateral signal at the 180$^{\circ}$ domain boundary (LS$_{\rm db}$) on BaTiO$_3$ is about 10\% when compared to the one on LiNbO$_3$ as it can be seen from the scanlines (c, f) which have the same arbitrary units.
The 'hot spots' indicated by the arrows in (b) originate from a subtle interplay between the different deformations of the crystal at the domain crossings.
The image size is $10 \times 5$\,\textmu m$^2$ each.%
}
\end{figure}

For another test of the model we carried out comparative measurements on multi-domain single-crystals of $\rm BaTiO_3$ and LiNbO$_3$. Figure~\ref{fig:Joh-03}(a,b) and (d,e) show the vertical and lateral PFM signals recorded simultaneously.
Obviously, the contrast-relation is the same for both materials: at a $\odot|\otimes$ domain boundary (e.g.~at the position of $\approx 6\,$\textmu m) the LS$_{\rm db}$ shows up as a bright peak; correspondingly at a $\otimes|\odot$ domain boundary, the LS$_{\rm db}$ shows up as a dark peak. As for the amplitudes of the LS$_{\rm db}$ is smaller by a factor of ten on $\rm BaTiO_3$ when compared to LiNbO$_3$.
\footnote{Subtracting the distorted baseline would increase the full amplitude of the LS$_{\rm bd}$ at most by a factor of two.}
The theoretical model, however, predicts a reversed contrast-relation for these two materials and a larger amplitude of the LS$_{\rm db}$ for BaTiO$_3$ than for LiNbO$_3$. Apparently, the model in its actual state can not explain our results.
\footnote{Improvements of the model might include the $d_{22}$ and also the effect of sideways clamping that suppresses shear deformations.}
%

\begin{figure}[hhh]
\includegraphics{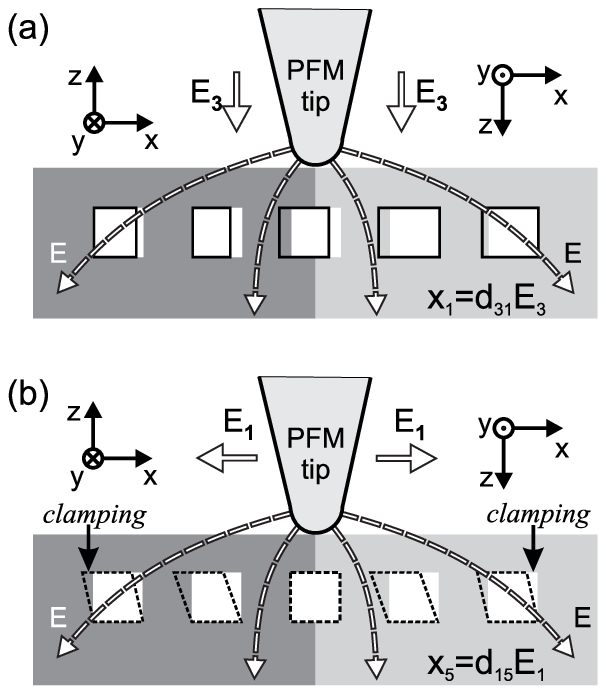}
\caption{\label{fig:Joh-04}
Two possible driving mechanisms for the lateral PFM signal at 180$^{\circ}$ domain walls owing to the electric field components $E_3$ perpendicular (a) and $E_1$ parallel (b) to the sample surface. The bright squares symbolize small volumes without an electric field applied. Their deformation when applying an electric field is shown in (a) by the solid contours for contraction/expansion and in (b) by the dotted contours for shearing. The latter deformation, however, is suppressed due to sideways clamping.
}
\end{figure}

We therefore attempted a qualitative understanding of the situation at the domain boundary.  At first, we would like to discuss the origin of lateral signals in PFM in a more general way. The electric field emerging from the tip can be approximated by that of a point charge. As a consequence, the sample will be deformed only in an "active volume" of few \textmu$\rm m^3$ size.\cite{Jun08a} Further out of this volume, the field strength is negligible, the sample will remain at rest. For lateral displacements of the sample surface one can now think about two mechanisms: expansion/contraction and shearing.\cite{Jun09} In the case of expansion/contraction the parts on the left side of the tip undergo a different deformation than those on its right side.(Fig.~\ref{fig:Joh-04}(a)). In the case of shearing, every part within the active volume undergoes the same deformation in one direction only  (Fig.~\ref{fig:Joh-04}(b)). Consequently, shearing is suppressed due to sideways clamping.
\footnote{A detailed description of this issue can be found in [\onlinecite{Jun09}] where in particular Fig.~5 is of interest.}

As a next step, we analyzed all possible combinations of a specific direction of the electric field~$E_i$ with the appropriate piezoelectric tensor elements~$d_{ij}$, the tip being placed on top of a 180$^{\circ}$ domain boundary. We thereby identified the $(E_i,d_{ij})$-pairs for $\rm LiNbO_3$, $\rm KTiOPO_4$, and $\rm BaTiO_3$ leading to an expansion/contraction displacement. We found that in the case of $\rm LiNbO_3$ (point group 3m) a  LS$_{\rm db}$ can be expected only on the $z$- and the $y$-faces but not on the $x$-face. For $\rm KTiOPO_4$ (point group mm2), only the polar $z$-face should show a LS$_{\rm db}$ whereas on the $y$-face no such signal is expected. Our experimental results match these predictions. Note that in these materials, for a given polarity of the tip, the direction of the lateral displacement at the domain boundary is the same for all $(E_i,d_{ij})$-pairs. The different contributions thus add up to a net LS$_{\rm db}$.

In the case of $\rm BaTiO_3$ (point group 4mm) the situation is somewhat more complicated. In this material, for a given polarity of the tip, the possible  $(E_i,d_{ij})$-pairs lead to lateral displacements at the domain boundary towards different directions,
thus partially canceling out. This might explain why on $\rm BaTiO_3$ only a very small LS$_{\rm db}$ was detected. A confirmation of this assumption might only be possible by means of finite element calculations.

In conclusion, we have discussed different mechanisms proposed so far to cause the lateral signal at the domain boundaries in PFM measurements. Although no final answer on the dominant driving mechanism can be given at the moment, it can, however, be stated that according to our experimental results: i)~slope-based explanations can be excluded, ii)~the electric field due to the domain specific surface charging can not solely cause the lateral signal at the domain boundaries and iii)~the actual theoretical model is not yet complete. We qualitatively discussed conceivable mechanisms for a lateral surface displacement at the domain boundary, the tip being placed on top of it. Finally we tried to explain why the lateral signal measured on BaTiO$_3$ is that small.

\vspace{1cm}

\footnotesize
{\bf Acknowledgments}
We thank P.~Paruch, J.~Guyonnet and H.~B\'{e}a  from the Universit\'{e} de Gen\`{e}ve for fruitful discussions and thereby launching our renewed investigation of that subject. We thank A.~N.~Morozovska from the Institute of Semiconductor Physics in Kiev for both, very open and extremely helpful discussions. We thank W.~Krieger, independent  gentleman, for fruitful discussions and careful reading of the manuscript. We thank D.~Rytz from FEE, Idar-Oberstein for providing the $\rm BaTiO_3$ sample. Financial support from the Deutsche Telekom AG is gratefully acknowledged.
\normalsize

\clearpage

\begin{center}
\bf Figure 1
\end{center}
\vspace{2cm}

\clearpage

\begin{center}
\bf Figure 2
\end{center}
\vspace{2cm}

\clearpage

\begin{center}
\bf Figure 3
\end{center}
\vspace{2cm}

\clearpage

\begin{center}
\bf Figure 4
\end{center}
\vspace{2cm}

\clearpage


\clearpage


\begin{thebibliography}{}


\bibitem{Gue92}
P. G{\"u}thner and K. Dransfeld,
Appl.\ Phys.\ Lett.\ \textbf{61}, 1137 (1992).

\bibitem{Alexe}
M.~Alexe and A.~Gruverman, eds., {\it Nanoscale Characterisation of
Ferroelectric Materials} (Springer, Berlin; New York, 2004) 1st ed.

\bibitem{Eng99}
 L. M. Eng, H.-J. G{\"u}ntherodt, G. A. Schneider, U. K{\"o}pke, and J. Mu{\~o}z Salda{\~n}a,
Appl. Phys. Lett. \textbf{74}, 233 (1999).

\bibitem{Kal06}
S.\,V.\,Kalinin, B.\,J.\,Rodriguez, S.\,Jesse, J.\,Shin, A.\,P.\,Baddorf,
P.\,Gupta, H.\,Jain, D.\,B.\,Williams, and A.\,Gruverman, Microscopy
and Microanalysis \textbf{12}, 206 (2006).

\bibitem{Jun09}
T.~Jungk, \'{A}.~Hoffmann, and E.~Soergel, New.\ J.\ Phys.\
\textbf{11}, 033029 (2009).

\bibitem{Gan02}
C.S. Ganpule, V. Nagarajan, B.K. Hill, A.L. Roytburd, E.D. Williams, R. Ramesh, S.P. Alpay, A. Roelofs, R. Waser, and L.M. Eng,
J. Appl. Phys. \textbf{91}, 1477 (2002).

\bibitem{Wit02}
J.~Wittborn, C.~Canalias, K.~V.~Rao, R.~Clemens, H.~Karlsson, and
F.~Laurell, Appl. Phys. Lett. \textbf{80}, 1622 (2002).

\bibitem{Scr05}
D.~A.~Scrymgeour and V.~Gopalan, Phys. Rev. B \textbf{72}, 024103
(2005).

\bibitem{Jun06b}
T.~Jungk, \'{A}.~Hoffmann, and E.~Soergel, Appl.\ Phys.\ Lett.\
\textbf{89}, 042901 (2006).

\bibitem{Guy09}
J.~Guyonnet, H.~Bea, F.~Guy, S.~Gariglio, S.~Fusil, K.~Bouzehouane, J.-M.~Triscone, and P.~Paruch,
Appl.\ Phys.\ Lett.\ \textbf{95}, 132902 (2009).

\bibitem{David}
D.~Scrymgeour, private communication: The $y$-axis of Fig.~14 in Ref.~[8] 
should have been labelled in [pm]. From this corrected graph one obtains $\alpha
=0.0062^{\circ}$.

\bibitem{Oka89}
H.~Okamura and J.~Minowa,
Electron.\ Lett.\ \textbf{25}, 395 (1989).

\bibitem{Col}
R.~J.~Colton, J.~E.~Frommer, A.~Engel, H.~E.~Gaub, R.~Guckenberger,
W.~M.~Heckl, B.~Parkinson, J.~Rabe, and A.~Gewirth, eds., {\it
Procedures in Scanning Probe Microscopies} (J.\ Wiley and Sons 1998).

\bibitem{Ter89} B.~D.~Terris, J.~E.~Stern, D.~Rugar, and H.~J.~Mamin,
 Phys.\ Rev.\ Lett.\ \textbf{63}, 2669 (1989).

\bibitem{Mor07}
A.~N.~Morozovska, E.~A.~Eliseev, S.~L.~Bravina, and S.~V.~Kalinin, Phys. Rev. B \textbf{75}, 174109 (2007).

\bibitem{Jun08a}
T.~Jungk, \'{A}.~Hoffmann, and E.~Soergel, New.\ J.\ Phys.\
\textbf{10}, 013019 (2008).



\end{thebibliography}
\end{document}